\documentclass{svjour2}     
\title{The role of time in relational quantum theories
}
\author{\bf Sean Gryb\footnote{\href{mailto:gryb0001@uu.nl}{gryb0001@uu.nl}}\\\it Institute for Theoretical Physics, Utrecht University\bigskip \\ \bf Karim Th\'ebault\footnote{\href{mailto:karim.thebault@gmail.com}{karim.thebault@gmail.com}}\\\it Centre for Time, Department of Philosophy, University of Sydney}

\usepackage{graphicx}
\usepackage{amssymb}
\usepackage{amsmath}
\usepackage[usenames,dvipsnames]{color}
\usepackage{hyperref}
\usepackage[utf8x]{inputenc}
\usepackage[left=1in,right=1in,top=1in,bottom=1in]{geometry}                  
\usepackage{epigraph}

\hypersetup{
    colorlinks=true,         
    linkcolor=blue,          
    citecolor=red,        
    urlcolor=Violet             
}

\let\oldmarginpar\marginpar
\renewcommand\marginpar[1]{\oldmarginpar{\color{red}\raggedright\scriptsize #1}}
\newcommand{\eq}[1]{(\ref{eq:#1})}
\newcommand{\pb}[2]{\ensuremath{\lf\{#1,#2 \rt\}}}

\newcommand{\diby}[2]{\ensuremath{\frac{\partial #1}{\partial #2}}}

\def\lf {\ensuremath{\left}}
\def\rt {\ensuremath{\right}}

\def\ham {\ensuremath{\mathcal H}}

\def\hpara {\ensuremath{\mathcal H_\Arrowvert}}
\def\hperp {\ensuremath{\mathcal H_\bot}}
\def\tint {\ensuremath{\tau_\text{int}}}
\def\teph {\ensuremath{\tau_\text{eph}}}
\def\ep {\ensuremath{\varepsilon}}

\begin{document}

\maketitle
\begin{abstract}
We propose a solution to the problem of time for systems with a single global Hamiltonian constraint. Our solution stems from the observation that, for these theories, conventional gauge theory methods fail to capture the \textit{full} classical dynamics of the system and must therefore be deemed inappropriate. We propose a new strategy for consistently quantizing systems with a relational notion of time that does capture the full classical dynamics of the system and allows for evolution parametrized by an \emph{equitable} internal clock. This proposal contains the minimal temporal structure necessary to retain the ordering of events required to describe classical evolution. In the context of \emph{shape dynamics} (an equivalent formulation of general relativity that is locally scale invariant and free of the local problem of time) our proposal can be shown to constitute a natural methodology for describing dynamical evolution in quantum gravity and to lead to a quantum theory analogous to the Dirac quantization of unimodular gravity.
\end{abstract}
\newpage
\tableofcontents

\section{Introduction}

\textit{Come what come may\\
Time and the hour runs through the roughest day.}\\
\begin{flushright}
--Macbeth, Act 1 Scene 3\\
\end{flushright}

\noindent
\newline As you read these words, is something happening? As your eyes scan across the page and your brain registers the meaning of the symbols, has something changed? Is there something physically different between this moment and the last? Every intuition we have about the nature of reality forces us to assert that \textit{something} must be happening. If this is so, then how can it have become common practice in canonical quantum gravity to assert that the evolution of our  universe is nothing but an unphysical gauge transformation? How can we reconcile ourselves to a physics locked in a static universe when our reality abounds with change?

There are two ways out of this conceptual \textit{cul--de--sac}. First, we could accept the fundamental timelessness of quantum gravity as described by the Wheeler--DeWitt equation (or something like it) and focus our energy on recovering the impression of dynamics from the frozen formalism. Much work has been devoted to this first path over the last twenty years: for example, the time capsules approach of Barbour \cite{barbour:eot,BarbourII,Halliwell:emergent_time}, the complete and partial observables scheme of Rovelli \cite{Rovelli:1990,Rovelli:1991,Rovelli:2002,Dittrich:2006,Dittrich:2007}, and approaches that rely on decoherence to regain time in a semi--classical approximation \cite{Kiefer:1988,Halliwell:1991,Halliwell:life_in_ee,Halliwell:2009}.\footnote{For an analysis of approaches broadly constituted along these lines, we refer the reader to the recent reviews of Anderson \cite{Anderson:review_pot,Anderson:2011}. For a comprehensive modern text on canonical quantum gravity see \cite{Thiemann:2007}. The  canonical formulation of general relativity was originally developed by Dirac \cite{Dirac:1958b} and  Arnowitt, Deser and Misner (`ADM') \cite{Arnowitt:1960,Arnowitt}.} 

The second escape route is to insist that it is our formalism rather than our intuition that has mislead us.\footnote{Defence of this largely unexplored second option is, to our knowledge, only found within the work of Kucha\v r \cite{kuchar:1991,kuchar:1992,Kuchar:1993,Kuchar:1999}.} We will present arguments demonstrating that conventional gauge theory methods\footnote{By \emph{conventional gauge theory methods} we will understand: i) Dirac quantization, ii) reduced phase space quantization, and iii) Faddeev--Popov gauge fixing, as outline in Section~\ref{sec:gf stand}.} \emph{can not} provide a quantization of the appropriate classical system when they are rigidly applied to theories where the dynamical trajectories are generated by a Hamiltonian constraint. Then, we will present an alternative that \emph{does} provide the appropriate quantization. In our procedure for \textit{relational quantization}\footnote{N.B. What we are proposing is only claimed to be a consistent methodology for the quantization of what is, under our definition, a relational theory. There are, of course, alternative types of quantization, just as there are alternative definitions of relationalism.} the minimal temporal structure of a physical theory is retained through the existence of a monotonically increasing parameter that labels the ordering of successive states. In contrast to a quantum theory that retains a Newtonian notion of absolute duration, in our proposal, duration is based upon relative change within the universe \emph{as a whole}. Our minimal notion of time is neither a pre--relativistic background parameter nor a quantity derived from an isolated subsystem. Central to our argument is the realisation that it is physically unreasonable to identify all points along a dynamical solution as a gauge orbit containing only physically equivalent states. Identifying all such points using a standard gauge fixing destroys the physical information about the temporal ordering of events and is inconsistent with the classical interpretation of reparametrization invariant theories. 

Many of the problems with regard to time and the quantization of general relativity originate in the curiously multifaceted -- part dynamical, part symmetry generating -- role that the local Hamiltonian constraints play within the classical canonical formalism. Individually, these constraints are responsible for foliation invariance and the associated local problem of time (see \cite{Thebault:ham_constraints} for discussion of the problem of interpreting local Hamiltonian constraints quantum mechanically). Collectively, they are responsible for reparametrization invariance and the associated global aspects of the problem of time. There exists, however, an equivalent formulation of general relativity, called \emph{shape dynamics}, that is free of the local problem of time. In this approach, developed in \cite{gryb:shape_dyn,Gomes:linking_paper}, there is a single Hamiltonian constraint generating the dynamics of conformal 3--geometry. The resulting global reparametrization invariance is the only remaining trace of the problem of time. In this paper, we will address, in particular, this \textit{global} aspect of the \textit{problem of time} through the consideration of analogue models and propose a quantization procedure that correctly captures the \emph{full} structure of the classical dynamics by retaining the ordering information of events in time. Then, we will study the gravitational case by applying our new quantization procedure to shape dynamics and general relativity. In both cases, we obtain a theory that is physically equivalent to the Dirac quantization of unimodular gravity. In shape dynamics, our proposal has a particularly clean implementation which resolves both \emph{local} and \emph{global} aspects of the problem of time.

\subsection{Main arguments}

In his influential \emph{Lectures on Quantum Mechanics} \cite{dirac:lectures}, Dirac claimed that all primary first class constraints are generators of infinitesimal transformations that do not change the physical state (p. 21). Despite his proof, it has been argued {\cite{kuchar:1991,kuchar:1992,Kuchar:1993,barbour:timelessness} that this assertion is invalid for first class constraints generating dynamical evolution. Furthermore, well--known counter examples to the conjecture are known to exist (see, for example, Section~1.2.2 of \cite{Henneaux:1992a}). As shown by Barbour and Foster \cite{barbour_foster:dirac_thm}, Dirac's proof operates under the assumption that the theory in question contains a fixed time parameter. Reparametrization invariant theories violate this assumption. Therefore, for this class of theories, Dirac's proof is not applicable and our interpretation of the constraints must be dictated by their physical origin. The potential relevance of this minority Kucha\v{r}-Barbour-Foster viewpoint to the \textit{diagnosis} of the problem of time in quantum gravity is clear: If it is not necessarily the case that Hamiltonian constraints must be treated as gauge generating, then one need not accept the conventional wisdom \cite{Henneaux:1992a} that, in reparametrization invariant theories, `motion is the unfolding of a gauge transformation' (p.104). In that context, one then has conceptual space to understand the derivation of a timeless Wheeler--DeWitt--type equation as the consequence of being lead to an inadequate quantum formalism by the application of an inappropriate interpretation of the classical constraints.  

The logical starting point of this paper is that the classical Hamiltonian constraints that feature in globally reparametrization invariant theories \textit{must} be treated as generating physical transformations associated with time evolution. In this context, we can establish that conventional canonical methodologies for dealing with classical gauge theories are, as they stand, inapplicable. Since the integral curves of the null directions associated with Hamiltonian constraints are solutions rather than equivalence classes of identical instantaneous states, what we would normally call `gauge orbits' are sequences of dynamically ordered physically distinct states. The ordering information is encoded in the positivity of the \emph{lapse} multiplier associated with the constraint. Contrary to what is found in the Hamiltonian formulation of all most all other gauge theories, the passage to a reduced phase space, where the null directions of the Hamiltonian constraints are quotiented out, leads to an initial data space lacking sufficient structure to reconstruct dynamics. This is because, to construct a solution from an initial data point, it is essential to retain the information of the ordering of events that is projected out by the reduction procedure. Formally, this is due to the fact that the pull back of the projection to the reduced space does not tell us how to temporally label the corresponding curves in the unreduced space.\footnote{We are indebted to Tim Koslowski for helping to clarify this key point.} We call this the \emph{problem of reduction}. As a result of this problem, a reduced phase space quantization will not lead to a quantum theory that correctly parametrizes the \textit{full} structure of the classical dynamics. Since such a quantization is, for the class of reparametrization invariant theories in question, equivalent to a gauge fixed path integral quantization \cite{Faddeev:1969} and a Dirac constraint quantization \cite{Gotay:1986}, the problem of reduction besets all three of these standard quantization techniques.  The problem of reduction will be illustrated in detail in Section~\ref{sec:prob_red} using path integrals for the quantum theory (Section~\ref{sec:rep inv}) and the Hamilton--Jacobi formalism for the classical theory (Section~\ref{sec:g vs d classical}).

In addition to these arguments stemming from the problem of reduction, there is good cause to re--evaluate the standard canonical quantization techniques on the grounds of an additional \textit{problem of relational time}. Classically, reparametrization invariant theories can be equipped with an internal \emph{and} equitable\footnote{Here, and below, by equitable we will mean dependent  upon the contributions of all the dynamical variables/subsystems of the system in question.} measure of duration that constitutes a fundamentally relational notion of time. Given strong conceptual and epistemological arguments in favour of physical models with relational time (see Section~\ref{sec: rel clocks}), we would like to construct a quantum theory with relational dynamics. However, as shown in Section~\ref{sec:int clock}, standard gauge theory techniques exclude the use of clocks that are both internal \emph{and} equitable (i.e. they are not relational by our definition). Thus, these techniques prohibit us from constructing a relation notion of time at the quantum level. Specific examples illustrating the problems of reduction and relational time are given in Section~\ref{sec:Jacobi}. 

We will thus, in Sections~\ref{sec:prob_red}-\ref{sec:Jacobi}, have provided ample arguments and evidence in favour of our unconventional diagnosis of the problem of time as owing to a failure of the formalism, rather than the timeless structure of the theories in question. In Section~\ref{sec:right way}, we will offer our proposal for a solution. Section~\ref{sec:formal procedure} will detail a formal procedure for retaining the essential dynamical ordering information through the introduction of an auxiliary field, and its momentum, that labels the classical trajectories and defines a relational time. The introduction of these variables is achieved via the extension of the phase space of the original theory in a precise geometrical manner.\footnote{We are again indebted to Tim Koslowski for his valuable insight in regards to this extension procedure.} We then show that the application of standard quantization techniques to the extended theory will lead to a quantum theory that correctly captures the full classical dynamics of the original theory we started with. Furthermore, as shown in Section~\ref{sec:solution}, the quantization of the extended theory is such that it leads to a quantum dynamics with respect to a relational time. This time is relational in the sense that it is a reparametrization invariant label for sequences of states and reduces to an equitable internal measure of duration in the classical limit. Our solution is applicable to all theories which feature a single global Hamiltonian constraints. It is not, therefore, immediately applicable to the full theory of general relativity, that features an infinite set of local Hamiltonian constraints. However, as detailed in Section~\ref{sec:gravSD}, when considered in the context of the \textit{shape dynamics} approach, gravity \textit{can} be understood in terms of a canonical formalism with a single global Hamiltonian constraint. We show that, when our quantization procedure is applied to shape dynamics, one obtains a quantum theory that is analogous to the Dirac quantization of unimodular gravity, where the cosmological constant is treated as a phase space operator canonically conjugate to the global relational time parameter. Thus, our relational quantization constitutes a proposal for solving the full problem of time in quantum gravity. 

As a precursor to our analysis of the problem of time, in the following section we review the key elements of Hamiltonian gauge theory that are necessary to frame our argument in precise terminology.

\section{Review of Hamiltonian gauge theory}\label{sec:gf stand}

In this section we review the essential structures of Hamiltonian gauge theory. Further details can be found in \cite{Arnold:1988,Souriau:1997}.

A Hamiltonian theory consists of at least three structures: 1) an even dimensional phase space $\Gamma(q_\mu, p^\mu)$ coordinatized by an equal number of configuration variables $q_\mu$ and their conjugate momenta $p^\mu$, where $\mu$ runs from 1 to the dimension $d$ of the system, 2) a closed, non--degenerate symplectic 2--form $\Omega = dq\wedge dp$, and 3) a Hamiltonian $H(q,p)$, which is a function on $\Gamma$. The non--degeneracy of $\Omega$ implies that its inverse, with coordinates $\Omega^{ab}$ in some chart (where $a,b$ range from 1 to $2d$), exists such that
\begin{equation}
    \Omega^{ac}\Omega_{cb} = \delta^a_b.
\end{equation}
Using $\Omega^{ab}$, we can define the Poison bracket between functions $f$ and $g$ on $\Gamma$ through the relation
\begin{equation}
    \pb{f}{g} = \Omega^{ab} \partial_a f \partial_b g
\end{equation}
where the derivatives are taken with respect to the coordinates on $\Gamma$. We can also define the Hamilton vector fields, $v_f$, of $f$ as
\begin{equation}
    v_f = \pb{f}{\cdot}.
\end{equation}
These define the tangent space of $\Gamma$. It is conventional to choose a representation for $\Omega^{ab}$ such that
\begin{equation}
    \Omega^{ab} = \lf( \begin{array}{c c}
		    0 & \delta^{\mu\nu} \\
		    -\delta^{\mu\nu} & 0 \\
                  \end{array} \rt).
\end{equation}
All structures are then defined up to symplectomorphisms, or \emph{canonical transformations}, which are diffeomorphisms of $\Gamma$ that preserve $\Omega$. The classical solutions of the theory are the integral curves of the Hamilton vector fields, $v_H$, of the Hamiltonian $H(q,p)$.

A Hamiltonian gauge theory has additional structure due the presence of first class constraints $\chi_i \approx 0$, where $i = 1,..,n$, that must be satisfied by the classical and quantum solutions.\footnote{Second class constraints can always be made first class by a suitable redefinition of $\Omega$ following Dirac \cite{dirac:lectures}.} These must form a first class algebra with themselves and $H$ under the action of the Poison bracket:
\begin{align}
    \pb{\chi_i}{\chi_j} &= c_{ij}^k \chi_k \\
    \pb{H}{\chi_i} &= k^j_i \chi_j,
\end{align}
for some structure constants $c_{ij}^k$ and $k^j_i$.

One immediately encounters a problem when trying to define the classical dynamics on the constraint surface, $\Sigma$, defined by $\chi_i \approx 0$: the pullback, $\omega \equiv \lf.  \Omega \rt\rvert_\Sigma$, of the symplectic 2--form, $\Omega$, onto $\Sigma$ is degenerate and closed (and, therefore, pre--symplectic). It is straightforward to show that $\omega$ will have $n$ null directions given by the Hamilton vector fields, $v_{\chi_i}$, of the constraints $\chi_i$. Therefore, the Hamilton vector fields of the Hamiltonian on the constraint surface are not unique and the classical evolution is non--deterministic.

In conventional gauge theory, the solution to this problem is implied by the physical interpretation of the integral curves of $v_{\chi_i}$ that foliate $\Sigma$. Every point along an integral curve is identified as a \emph{physically indistinguishable} state of the system. Consequently, the leaves of $\Sigma$ are interpreted as equivalence classes of physical states and are named \emph{gauge orbits}. One way to restore determinism would be to select a particular member of each gauge orbit and therefore select a unique representative for each physically distinguishable state. This surface constitutes a gauge--fixed, $2(d-n)$ dimensional submanifold, $\Sigma_\text{gf}$, of $\Gamma$. Furthermore, $\Sigma_\text{gf}$ is symplectic because the pullback, $\Omega_\text{gf} \equiv \lf.  \Omega \rt\rvert_{\Sigma_\text{gf}}$, of $\Omega$ on $\Sigma_\text{gf}$ is, by construction, non--degenerate. The symplectic manifold $(\Sigma_\text{gf}, \Omega_\text{gf})$ is equivalent to a reduced phase space constructed from the quotient of the constraint surface $\Sigma$ by the null space of $\omega$. Crucial to this reduced formalism is the assumption that the map $\pi: \Sigma \to \Sigma_\text{gf}$ encodes \emph{no} dynamical information. We will see in Section~\ref{sec:g vs d classical} that this assumption breaks down in theories with global Hamiltonian constraints.

The specification of $\Sigma_\text{gf}$ is achieved in practice by imposing gauge fixing conditions $\rho_i \approx 0$ such that the Poison bracket
\begin{equation}
    \pb{\chi_i(t)}{\rho_j(t')}
\end{equation}
seen as a matrix in $t$, $t'$, $i$, and $j$ is invertible. Equivalently,
\begin{equation}\label{eq:gf cond}
    \det \lf| \pb{\chi_i(t)}{\rho_j(t')} \rt| \neq 0.
\end{equation}
$\Sigma_\text{gf}$ is then defined as the non--degenerate intersection of $\chi_i \approx 0$ and $\rho_i \approx 0$. The condition \eq{gf cond} ensures that the gauge fixing only selects a single member of each gauge orbit. In other words, $\Sigma_\text{gf}$ must be nowhere parallel to $v_{\chi_i}$.\footnote{There is also a global requirement that $\Sigma_\text{gf}$ only intersects the gauge orbits once. We will assume that this requirement can be satisfied.} This geometric requirement of the gauge fixing will be important for our argument later.

Given a set of gauge fixing conditions $\rho_i\approx 0$ satisfying \eq{gf cond}, it is possible to define the classical and quantum theories. The classical solutions are the integral curves of the Hamiltonian on $\Sigma_\text{gf}$. The quantum theory is defined by integrating over all paths on the gauge fixed surface restricted by the endpoints $q_\text{in}$ and $q_\text{fin}$. The integration measure over the gauge fixed surface is given by the only non--trivial structure available to us from the formalism:
\begin{equation}
    \det \lf| \pb{\chi_i(t)}{\rho_j(t')} \rt|.
\end{equation}
This measure is precisely the Jacobian obtained from a canonical transformation from $\Gamma$ to the reduced phase space. Thus, the path integral
\begin{equation}
    I(q_\text{in},q_\text{fin}) = \int \lf. \mathcal Dq_\mu \rt\rvert_{q_\text{in},q_\text{fin}} \mathcal Dp^\mu \delta(\chi_i) \delta(\rho_i) \det \lf| \pb{\chi_i(t)}{\rho_j(t')} \rt| \exp \lf\{ i \int dt \lf[ \dot q_\mu p^\mu - H \rt] \rt\}
\end{equation}
is equal to a Feynman path integral over the reduced phase space. For details, see \cite{faddeev:fp}. Additionally, it's been conjectured \cite{Guillemin:1982}, and proven in limited cases (e.g., \cite{Gotay:1986}), that quantization of the reduced phase space is equivalent Dirac quantization of constrained systems \cite{dirac:lectures}. We will in the following discussion group the three quantization techniques together as the \textit{standard} or \textit{conventional} methodologies for the quantization of a gauge theory. 

\section{Diagnosing the problem of time i: the problem of reduction}\label{sec:prob_red}

\subsection{Gauge invariance versus dynamics: quantum}\label{sec:rep inv}

Globally reparametrization invariant theories feature action functionals in which the integration is performed with respect to an \textit{arbitrary change parameter} $\lambda$ rather than a fixed Newtonian background time. The invariance of these theories under re--scalings of this parameter leads to the defining feature of their canonical representation: that the Hamiltonian is  replaced with a constraint $\ham$, often called the \emph{Hamiltonian constraint}. Assuming that all other first class constraints have been gauge--fixed using the method described above,\footnote{Crucially, this is the step that can be performed in shape dynamics that is highly non--trivial in the ADM formulation of general relativity.} the remaining structure is a phase space $\Gamma(q,p)$ (possibly corresponding to a gauge fixed surface in a larger phase space), a symplectic 2--form, $\Omega$, and a Hamiltonian constraint, $\ham$.

We now state the fundamental difference between reparametrization invariant theories and standard gauge theories: the classical solutions are defined as the integral curves of the Hamilton vector field of the constraint $\ham$. Because they are the dynamical solutions of the classical theory, the elements of the integral curves of $v_\ham$ are no more \emph{physically indistinguishable} from each other than this moment is from the big bang. Thus, the leaves of the foliations of the constraint surface, $\Sigma$, defined by $\ham \approx 0$ can no longer be reasonably identified as gauge orbits - rather they are dynamical solutions. We will now show that, if one turns a blind eye to this fact, one is led to a quantum theory that, in general, cannot contain the appropriate classical limit.

Performing the gauge fixing procedure outlined in Section~\ref{sec:gf stand}, we treat each classical history on the constraint surface as an equivalence class of physically indistinguishable states. We then seek a gauge fixing condition $\rho \approx 0$ satisfying
\begin{equation}
    \det \lf| \pb{\ham}{\rho} \rt| \neq 0
\end{equation}
that selects a single element of each of these foliations. However, because the Hamiltonian is just given by the constraint $\ham$, its associated flow is everywhere parallel to the gauge orbits. Thus, this procedure completely trivializes the dynamics since there is no way to flow in any direction on the gauge fixed surface. In addition, the interpretation of the gauge fixed surface is now completely different from the case described in Section~\ref{sec:gf stand}. $\Sigma_\text{gf}$ contains a  single element of each integral curve of $v_\ham$. Since each integral curve is itself a possible classical solution, $\Sigma_\text{gf}$ actually represents a space of initial data for all possible classical evolutions on the constraint surface. Thus, the space, by construction, necessarily excludes any (non-trivial) set of points on the constraint surface corresponding to a classical history. We therefore have that the path integral
\begin{equation}\label{eq:pi bad}
    I = \int \mathcal Dq_\mu \mathcal Dp^\mu \delta(\ham) \delta(\rho) \det \lf| \pb{\ham}{\rho} \rt| \exp \lf\{ i \int d\lambda \lf[ \dot q_\mu p^\mu \rt] \rt\}
\end{equation}
restricted to $\Sigma_\text{gf}$ cannot contain any particular solution to the classical evolution problem. It, therefore, \emph{can not} be a quantization of the original classical theory. This is equivalent to the statement that the Feynman path integral on the reduced space, which is canonically isomorphic to $\Sigma_\text{gf}$, fails to capture the classical evolution. In Section~\ref{sec:jac ho}, we give an explicit example illustrating this point.


One key problem we will solve in this paper will be establishing a consistent quantization procedure for globally reparametrization invariant theories that \emph{does} contain the appropriate classical limit. We are faced with a dilemma: on one hand, we need to restrict our path integral to a proper symplectic manifold where the Hamilton vector field of $\ham$ is well defined on the constraint surface; but, on the other hand, such a restriction must be such that the constraint $\rho \approx 0$ runs parallel to the foliations of $\ham$. Unfortunately, this would imply
\begin{equation}
    \det \lf| \pb{\ham}{\rho} \rt| = 0
\end{equation}
and we no longer have a natural candidate for the measure of the path integral. The solution that we will propose in Section~\ref{sec:right way} involves extending the phase space in a trivial way so that the desired classical solutions are indeed contained in the initial value problem of the extended theory. Thus, a standard gauge fixing on this extended theory corresponds to a consistent quantization of the original theory. Before describing this procedure in detail, we will show how the argument presented above is paralleled in the classical theory.

\subsection{Gauge invariance versus dynamics: classical}\label{sec:g vs d classical}

In the semi--classical approximation, the wavefunction of a system is given by the WKB ansatz
\begin{equation}
    \psi = e^{iS},
\end{equation}
where $S$ solves the Hamilton--Jacobi (HJ) equation. When the dynamics is generated by a constraint, the HJ equation takes the form
\begin{equation}\label{eq:HJ main}
    H(q_\mu, \diby{S}{q_\mu}) = 0.
\end{equation}
Hamilton's principal function $S = S(q_\mu, P^a)$ is a function of the configuration variables, $q_\mu$, and the separation constants, $P^a$. These separation constants are obtained by solving the partial differential equation \eq{HJ main}. In general, there will be one for each $\diby S {q_\mu}$ but these will not all be independent because \eq{HJ main} acts as a constraint. This is the reason for labeling $P$ with the index $a$, which runs from $1$ to $d-1$.

The equations of motion are obtained by treating $S$ as a generating function for a canonical transformation from $(q_\mu, p^\mu) \to (Q_a, P^a)$ that trivializes the evolution. The canonical transformation can be determined by computing
\begin{align}
    Q_a &= \diby {S(q, P)}{P^a} \label{eq:Q}\\
    p^\mu &= \diby{S(q,P)}{q_\mu}\label{eq:p}.
\end{align}
$S$ is defined such that the relations \eq{p} simply reproduce the Hamiltonian constraint through \eq{HJ main}. If the system of equations \eq{Q} can be inverted for $q_\mu$ then the equations of motion for $q_\mu$ can be determined by using the fact that
\begin{align}
    \dot Q_a & = 0 & \dot P^a & = 0.
\end{align}

There is, however, an immediate obstruction to this procedure since the system of equations \eq{Q} has, in general, a one dimensional kernel and, thus, no unique solution. This obstruction can be overcome in two ways:
\begin{enumerate}
    \item A \emph{gauge} can be fixed by imposing a gauge fixing condition of the form
    \begin{equation}\label{eq:gf}
	f(q_\mu, Q_a, P^a,\lambda) = 0.
    \end{equation}
    $f$ must be chosen such that, when the condition $f = 0$ is imposed, the system of equations \eq{Q} is invertible.
    \item The solution space can be parametrized by one of the $q$'s, chosen arbitrarily. This allows us to write
    \begin{equation}\label{eq:int curve}
	q_a = F_a(Q_a, P^a, q_0).
    \end{equation}
\end{enumerate}

The first method is the one exclusively employed to conventional gauge theories (i.e. those without Hamiltonian constraints). The gauge fixing \eq{gf} reduces the dimension of the system. This is natural in standard gauge theory because the map between the original and reduced phase spaces contains no physical information. It is, therefore, reasonable to make the equations of motion invertible by quotienting away the information contained in this map. This does not kill the dynamical information because a non--trivial Hamiltonian survives the quotienting. However, for globally reparametrization invariant theories, the information contained in the kernel of \eq{Q} contains \emph{all} the dynamical information. Thus, we must use the second method for reproducing the classical solutions. This is natural, because the relations \eq{int curve} are precisely the integral curves of null directions of the presymplectic form on the constraint surface $H = 0$. In must be noted here that, with regard to this particular point, our analysis is not controversial: method 2 precisely coincides with Rovelli's \cite{Rovelli:2004} treatment of Newtonian particles (see pp.113-4) and is consistent with how HJ theory is used to reproduce the ADM equations of motion in general relativity \cite{Gerlach:EHJ}. The reader is referred to these sources, and references therein, for more details on this standard treatment. In Section~\ref{sec: Jac fp}, we will apply methods 1 and 2 to a simple model to illustrate how to implement the formal procedure presented here.

A powerful argument can be made in favour of method 2 over method 1, when dealing with reparametrization invariant theories. In method 1, the pullback under the projection doesn't contain the complete dynamical information. Only in method 2 is it possible to retain information about the temporal ordering of events along the gauge orbits. This a necessary requirement for theories with Hamiltonian constraints because there must be a way to distinguish between the past and the future.\footnote{This distinction constitutes a temporal orientation rather than a temporal direction, which would imply an arrow of time.} This is already implicit in requiring that the lapse, $N$, should be positive. The fact that \emph{only} Hamiltonian constraints have this requirement is an indication that they should be treated differently from the constraints arising in standard gauge theories.

We see that there is a substantive difference between the way the HJ formalism is used in conventional gauge theory and in globally reparametrization invariant theories. This difference is exactly mirrored in the quantum theory. The arguments given in Section~\ref{sec:rep inv} reflect what happens in the classical theory when method 1 is used: the information about the dynamics is lost by quotienting with respect to the null directions of the Hamiltonian flow. A requirement for consistency for the classical and quantum theories is that the method used in the classical theory is reflected in that used in the quantum theory. In light of this requirement and the necessity of using method 2 classically for reparametrization invariant theories, we will present a relational quantization procedure in Section~\ref{sec:right way}.

\section{Diagnosing the problem of time ii: the problem of relational time}

\subsection{Relational clocks}\label{sec: rel clocks}

There is, without doubt, practical utility in the use of a time parameter disconnected from the dynamics of a physical system. Such an external notion of time is an essential element of both Newtonian systems and conventional approaches to quantum theory. Yet, the existence of such a temporal background is inconsistent without the structure of the physical theory that most accurately describes the behaviour of clocks: general relativity. Within this theory, time is an inherently internal notion, parasitic upon the dynamics. Thus, there is empirical motivation to search for a general procedure for consistently constructing an internal notion of time that can be used in both classical and quantum theories.

In addition, there are strong conceptual arguments against external time -- many of which predate general relativity. Ernst Mach, in particular, criticized external notions of time on epistemic grounds. In the most general system, we only have access to the internal dynamical degrees of freedom. Thus, it is ``utterly beyond our power to measure the changes of things by time'' \cite{mach:mechanics}. Rather, according to Mach, any consistent notion of time must be abstracted from change such that the inherently interconnected nature of every possible internal measure of time is accounted for. According to the Mittelstaedt--Barbour \cite{Mittelstaedt:machs_2nd,barbour:newton_2_mach} interpretation of Mach, we can understand this \textit{second Mach's principle} as motivating a relational notion of time that is not merely internal but also equitable; in that it can be derived uniquely from the motions of the entire system taken together. Thus, any isolated system -- and, in fact, the universe as a whole -- would have its own natural clock emergent from the  dynamics. Significantly, for a notion of time to be relational in this sense, it is not enough to be merely internal -- it must also be unique and equitable. We cannot, therefore, merely identify an isolated subsystem as our relational clock, since to do so is not only non--unique but would also lead to an inequitable measure, insensitive to the dynamics of the clock system itself.

Within classical non--relativistic theory, relational clocks of exactly the desired type have already been constructed and utilized. As has been pointed out by Barbour, the astronomical measure of \textit{ephemeris time}, based upon the collective motions of the solar system, has precisely the properties discussed above. In Section~\ref{sec:Jacobi}, we give an explicit expression for the ephemeris time for a large class of physically relevant finite dimensional models. Quantum mechanically, we run into a problem when attempting to construct a suitably relational notion of time. As we shall discuss in the next section, it is precisely the relational sub--set of internal clocks that are excluded under conventional quantization techniques. The logic of the next section is as follows: first we establish a general theory for describing evolution in timeless systems in terms of an internal clock as constituted by an isolated subsystem; then, we show that such clocks can never be fully adequate precisely because they are not fully relational.

\subsection{Internal clocks}\label{sec:int clock}

We will now detail a method for expressing the path integral \eq{pi bad} in terms of evolution with respect to an internal clock constructed from an isolated subsystem. Consider \emph{any} splitting of the Hamiltonian constraint of the form:
\begin{equation}\label{eq:split}
    \ham = \hpara + \hperp.
\end{equation}
Our ability to make this splitting depends principally upon the existence of a sufficiently isolated clock. In practice, the split need only be approximate to some desired order of accuracy. Effectively, we require a clock of the form treated in great detail in \cite{Marolf:Internal_Time}. For more details on the use of internal clocks as a way of modeling relational dynamics and some of the difficulties encountered see \cite{Marolf:Mini_Superspace,Giddings:Effective_observables,Bojowald:effective_pot,Hilgevoord:2005}. Given that we have an approximate splitting of the form \eq{split}, we are in the situation treated in the above references and we can perform a canonical transformation $\Pi$
\begin{equation}
    \Pi: (q_\mu,p^\mu) \to (Q_i, P^i, \tint, \hpara)
\end{equation}
generated by the type--2 generating functional $F(q_\mu, P^a, \hpara)$

\begin{equation}
    F(q_\mu, P^a, \hpara) = \int dq_\mu\, p^\mu(q_\mu, P^a, \hpara).
\end{equation}
The index $i$ runs from $1,..,d-1$. The functions $p^\mu(q_\mu, P^a, \hpara)$ are obtained by inverting the relations
\begin{align}
    P^a &= P^a(q_\mu, p^\mu) \notag \\
    \hpara(q_\mu, p^\mu) &= \ham(q_\mu, p^\mu) - \hperp(q_\mu, p^\mu).\label{eq:relations}
\end{align}
The functions $P^a(q_\mu, p^\mu)$ are arbitrary provided the above equations are invertible for $p^\mu$. Because $\hpara$ is fixed by the splitting \eq{split}, the canonical transformation $\Pi$ has a $(d-1)$--parameter freedom parametrized by the functions $P^a(q_\mu, p^\mu)$. Up to this freedom, $\Pi$ singles out an \emph{internal time} variable $\tint$ which can be obtained from
\begin{equation}
    \tint (q_\mu, p^\mu)= \lf. \diby{F}{\hpara} \rt\rvert_{P^a = P^a(q_\mu, p^\mu), \hpara = \hpara(q_\mu, p^\mu)}.
\end{equation}
We say that $\hpara$ singles out an isolated subsystem of the universe whose motion is used as an internal clock parametrizing the motion of the rest of the system. The remaining configuration variables are given by
\begin{equation}
    Q_i(q_\mu, p^\mu) = \lf. \diby{F}{P^i}\rt\rvert_{P^a = P^a(q_\mu, p^\mu), \hpara = \hpara(q_\mu, p^\mu)}.
\end{equation}

If we chose to label curves in $\Gamma$ by the arbitrary parameter $\lambda$ then, in terms of the transformed coordinates, the natural gauge fixing condition
\begin{equation}
    \rho = \tint - f(\lambda) \approx 0
\end{equation}
splits $\Gamma$ into the two symplectic submanifolds $\Gamma = \Sigma_\text{gf}(Q_a, P^a) \times \Phi(\tint,\hpara)$. From this, it is clear that the freedom in defining $\tint$ through  $\Pi$ corresponds to the unavoidable freedom in making arbitrary canonical transformations on $\Sigma_\text{gf}$. Using the simple choice $f(\lambda) = \lambda$, the measure is 1. A short calculation shows that the path integral, $I$, of \eq{pi bad} becomes
\begin{equation}\label{eq:emergent t}
    I = \int \mathcal DQ_i \mathcal DP^i \exp \lf\{ i \int d\tint \lf[ \dot Q_a P^a - \hperp(Q_a,P^a) \rt] \rt\}
\end{equation}
after integrating over $\tint$ and $\hpara$. As expected, it is equivalent to a Feynman path integral on the reduced phase space coordinatized by $(Q_a,P^a)$. This path integral also corresponds to the quantization of a standard time--dependent Hamiltonian theory in terms of the variables $Q^a$ and momenta $P^a$ with the Hamiltonian $H = \hperp (Q_a,P^a)$. It leads to a wavefunction satisfying the time--dependent Schr\"odinger equation
\begin{equation}
    i\diby{\Phi}{\tint} = \hat\hperp \Psi.
\end{equation}
Furthermore, this is equivalent to applying Dirac quantization to the Hamiltonian constraint \eq{split} after applying $\Pi$.\footnote{See \cite{Henneaux:1992a} p.280 for a analogous case.}

The quantum theory given by \eq{emergent t} has a well known classical limit. It is the Hamiltonian theory described by the integral curves of $\hperp$ parametrized by $\tint$. Different choices of $f(\lambda)$ correspond to different parametrizations of these integral curves. Although this freedom to reparametrize the classical solutions is a feature we require, the classical solutions we obtain are \emph{not} equal (or equivalent) to the integral curves of $\ham$. Instead they are the integral curves of the part projected out of $v_\ham$ along $\Phi(\tint, \hpara)$. Thus, the only way to obtain the desired classical limit is to impose $\hpara = 0$.  \textit{However, with this choice the  above method fails since the relations \eq{relations} become non--invertible}. This is another way of understanding the problem of reduction: gauge fixing such that we follow the integral curves of $v_\ham$ leads to a zero measure in the path integral. In addition to the problem of excluding classical trajectories, this restriction on internal clocks is such that it specifically excludes relational clocks of the kind considered in the previous section. In phase space, the classical ephemeris time is precisely the variable canonically conjugate to the full Hamiltonian of the system. In the following section we demonstrate this explicitly in a simple model.

\section{Toy models}\label{sec:Jacobi}

We will now make our formal arguments concrete by applying them to specific models. These models will also help to motivate our new quantization procedure. We will consider models of the form
\begin{equation}\label{eq:act jac}
    S = \int d\lambda \sqrt{ g^{\mu\nu}(q) \dot q_\mu \dot q_\nu },
\end{equation}
where $g$ is some specified metric on configuration space. The variation of this action with respect to $q$ implies that it is a geodesic principle on configuration space. Thus, \eq{act jac} is invariant under $\lambda \to f(\lambda)$, which is the reparametrization invariance we require. These models are useful gravitational models because they include the mini--superspace approximation and contain many key features of general relativity and shape dynamics (See Section~\ref{sec:gravSD}). The fact that these simple models can capture so many features of gravity is often under-appreciated. Indeed, because they correspond to mini--superspace models they are, in fact, genuine symmetry reduced models of quantum gravity. Furthermore, because they are also equivalent to non--relativistic particle models, they have considerable heuristic value.

We will treat the case where $g_{ab}$ is conformally flat. Identifying the conformal factor with $2(E - V(q))$, the Hamiltonian constraint is
\begin{equation}\label{eq:ham jac}
    \ham = \frac{\delta_{\mu\nu}}{2} p^\mu p^\nu + V(q) - E \approx 0.
\end{equation}
It is important to note that the origin of this constraint can be traced back to the reparametrization invariance of the action \eq{act jac}. As such, its interpretation is crucially different from that of the gauge generating constraints discussed in Section~\ref{sec:gf stand}.

The classical theory corresponding to the Hamiltonian \eq{ham jac} is just that of non--relativistic particles under the influence of a potential $V(q)$ with total energy $E$ and mass $m=1$. The classical equations of motion are easily seen to lead to
\begin{equation}\label{eq:jac eom}
    \sqrt{ \frac{E-V}{T}} \frac{d}{d\lambda} \lf( \sqrt{\frac{E-V}{T}} \frac{dq_\mu}{d\lambda} \rt) = - \diby{V}{q^\mu},
\end{equation}
where $T = \frac{\delta_{\mu\nu}}{2} p^\mu p^\nu$ is the kinetic energy. If we define the reparametrization invariant quantity
\begin{equation}
    \teph \equiv \int d\lambda \sqrt{\frac{T}{E - V}}
\end{equation}
then \eq{jac eom} becomes
\begin{equation}
    \frac{d^2 q_\mu}{d\tau^2} = -\diby{V}{q^\mu},
\end{equation}
which are Newton's equations with $\teph$ playing the role of absolute time. Newton's theory is then given by the integral curves of \eq{ham jac} parametrized by the ephemeris time label $\teph$.

\subsection{Example: double pendulum}\label{sec:jac ho}

Consider the double pendulum consisting of 2 particles $q_1$ and $q_2$ in 1 dimension under the influence of a potential
\begin{equation}
    V(q) = \frac 1 2 \lf( q_1^2 + q_2^2 \rt)
\end{equation}
corresponding to 2 uncoupled harmonic oscillators whose spring constants $k$ have been set to 1. The Hamiltonian constraint is\footnote{In this section we will sometimes write the coordinates of $p$ using lower case indices for convenience.}
\begin{equation}
    \ham_\text{HO} = \frac 1 2 \lf( p_1^2 + p_2^2 + q_1^2 + q_2^2\rt) - E.
\end{equation}
Its Hamilton vector field $v_\ham$ is
\begin{equation}
    v_\ham = p^\mu \diby{}{q_\mu} - q_\mu \diby{}{p^\mu}
\end{equation}
and the constraint surface is the $S^3$ boundary of the 4--sphere of radius $E$. The integral curves on $\ham = 0$ are circles when projected into the $(q_\mu, p^\mu)$--planes as is familiar from the usual harmonic oscillator.

Performing a standard gauge fixing, as described in Section~\ref{sec:gf stand}, and following the procedure described in Section~\ref{sec:int clock}, we split the Hamiltonian constraint into the pieces
\begin{align}
    \hpara &= \frac 1 2 \lf( p_1^2 + q_1^2 \rt) & \hperp &= \frac 1 2 \lf( p_2^2 + q_2^2 \rt) - E.
\end{align}
Using this splitting, we can single out particle 1 as an internal clock for the system. We perform a canonical transformation that takes us to the internal clock variables for particle 1 and leaves particle 2 unchanged. The relations \eq{relations} become
\begin{align}
    p_2 &= P_2 \\
    \hpara &= \frac 1 2 \lf( p_1^2 + q_1^2 \rt).
\end{align}
Inverting these, we are led to the generating functional
\begin{equation}
    F = \int dq_1 \sqrt{2\hpara - q_1^2} + q_2 P_2.
\end{equation}
The transformed $Q_2$ coordinate is $q_2$ as expected and the internal time variable canonically conjugate to $\hpara$ is
\begin{equation}
    \tint = \lf. \diby{F}{\hpara} \rt\rvert_{\hpara = \frac 1 2 \lf( p_1^2 + q_1^2 \rt)} = \arctan \lf( \frac{q_1}{p_1} \rt).
\end{equation}
As can be seen from the definitions of $\tint$ and $\hpara$ in terms of $q_1$ and $p_1$, this canonical transformation takes us to polar coordinates on the $(q_1,p_1)$--plane of phase space.

The transformed Hamiltonian is
\begin{equation}
    \ham = \hpara + \frac 1 2 \lf( P_2^2 + Q_2^2 \rt) - E.
\end{equation}
Its Hamilton vector field is
\begin{equation}
    v_\ham = P_2 \diby{}{Q_2} - Q_2 \diby{}{P_2} + \diby{}{\tint}.
\end{equation}
The constraint surface is a cylinder along the $\tint$ direction about the $(q_2,p_2)$--plane of radius $E-\hpara$. The integral curves of $v_\ham$ are helices along the $\tint$--direction and wrap around the $\hpara$--direction implying that $\hpara$ is a classical constant of motion (see Figure (\ref{fig:int curves sho})).
\begin{figure}
 \centering
 \includegraphics[width=.6\textwidth]{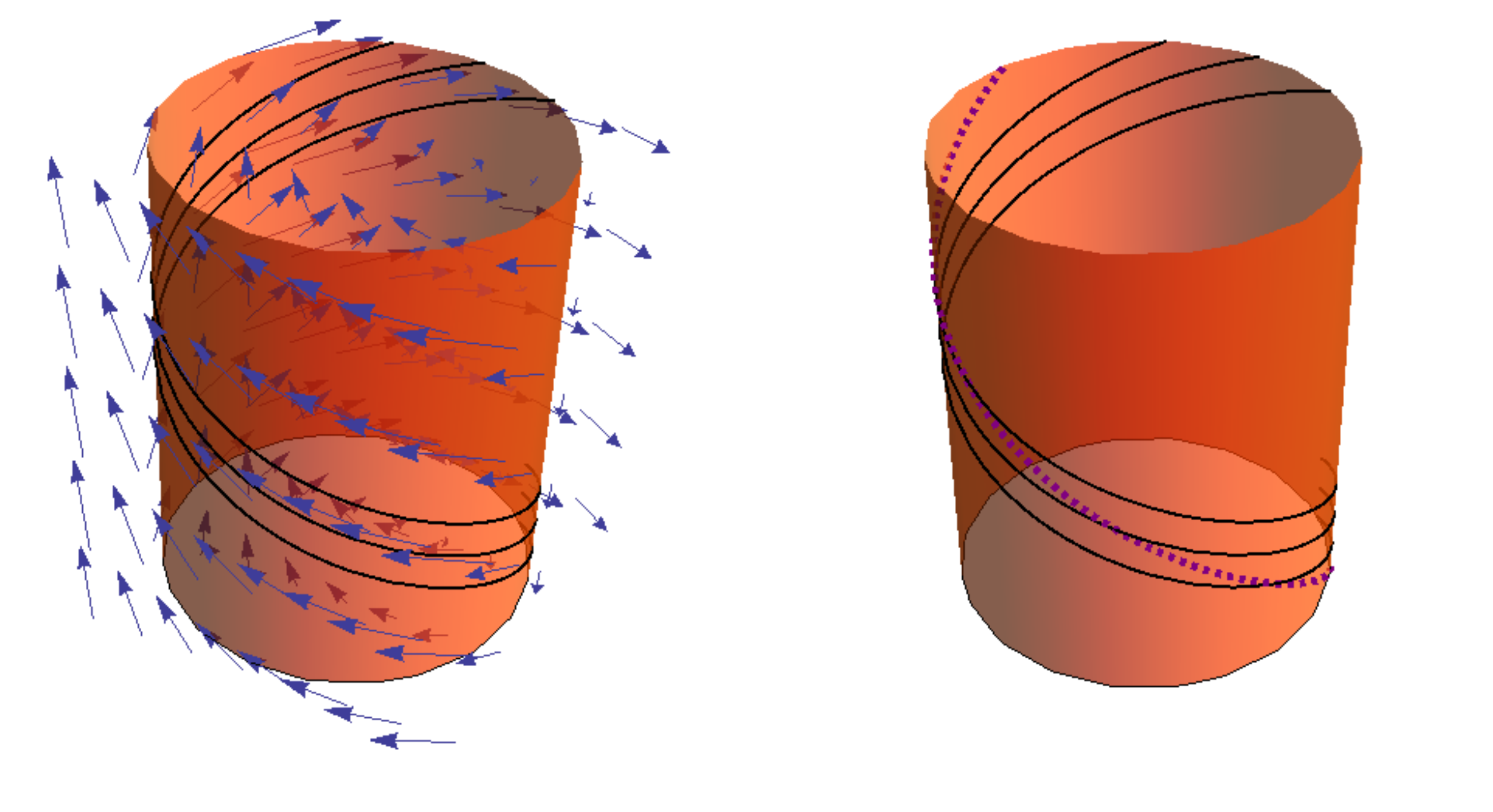}
 \caption{\small{The left hand graphic shows the constraint surface $\hpara=0$, the vector field $v_\ham$ and three examples of classical solutions (integral curves of $v_\ham$). The right hand graphic shows, as a dashed line, a sample path that is included in the integral \eq{SHO_pathint}, but, by definition, is nowhere parallel to the classical solutions.}}
 \label{fig:int curves sho}
\end{figure}

If we impose the gauge fixing condition $\tint = \lambda$, the path integral \eq{emergent t} takes the form
\begin{equation}\label{eq:SHO_pathint}
    I_\text{HO} = \int \mathcal DQ_2 \mathcal DP_2 \exp \lf\{ i \int d\tint \lf[ \dot Q_2 P_2 - \frac 1 2 \lf( P_2^2 + Q_2^2 \rt) + E \rt] \rt\},
\end{equation}
which leads to the time--dependent Schr\"odinger equation
\begin{equation}
    i \diby{\Psi}{\tint} = \lf( -\frac 1 2 \diby{^2}{Q_2^2} + \frac 1 2 Q_2^2  - E \rt) \Psi.
\end{equation}
This is the same theory we would have obtained had we quantized the 1D harmonic oscillator in the usual way. However, the freedom in redefining $\tint = f(\lambda)$ allows us the freedom to reparametrize the paths in phase space arbitrarily. Although this is the reparametrization freedom we want, it doesn't give the freedom to reparametrize the \emph{full} set of classical solutions.

An easy way to see that these paths will be excluded is to realize that these paths will contribute to the path integral with zero measure because the determinant $\det \lf| \pb{\ham}{\tint} \rt|$ is zero for these paths. On the other hand, the paths that \emph{are} captured in the integration are the ones corresponding to the 1D harmonic oscillator when projected down to the $(q_2,p_2)$--plane. This fact is reflected in our final result. In other words, this gauge fixing has effectively quantized the reparametrization invariant 1D harmonic oscillator, \emph{not} the 2D oscillator we started with.

\subsection{Example: relational free particle}\label{sec: Jac fp}

In this section, we will solve for the classical trajectories of the relational free particle using the HJ formalism. We will compare methods 1 and 2, presented in Section~\ref{sec:g vs d classical}, to show why conventional gauge theory methods should not be used in this case.

For the free particle, the Hamiltonian constraint \eq{ham jac} takes the form
\begin{equation}
    \ham = \frac{\delta_{\mu\nu}}{2} p^\mu p^\nu - E \approx 0.\label{eq:fp ham}
\end{equation}
Thus, the HJ equation reads
\begin{equation}
    \frac {\delta_{\mu\nu}} 2 \diby S {q_\mu} \diby S {q_\nu} - E = 0.
\end{equation}
This equation can be explicitly solved by introducing the $d-1$ separation constants $P^a$. The solution is
\begin{equation}
    S(q_\mu, P^a) = q_a P^a \pm \sqrt{2E - P^2} q_0,
\end{equation}
where $P^2 = \delta_{ab} P^a P^b$. We can solve the classical equations of motion by solving for $Q_a$ and $p^\mu$, then by inverting these relations in terms of $q_\mu$. Differentiating $S$ gives
\begin{align}
    Q_a &= \diby S {P^a} = q_a \mp \frac{ \delta_{ab} P^b } { \sqrt{ 2E - P^2 } } q_0 \label{eq:fp Q}\\
    p^a &= \diby S {q_a} = P^a \\
    p^0 &= \diby S {q_0} = \pm \sqrt{ 2E - P^2 }.
\end{align}
We recover the Hamiltonian constraint, \eq{fp ham}, immediately from the last two relations for $p^\mu$. As expected, \eq{fp Q} is non--invertible for $q_\mu$.

There are two possible ways to deal with the non--invertibility of \eq{fp Q}:
\begin{itemize}
    \item {\bf Method 1:} Impose the gauge fixing condition
    \begin{equation}
	q_0 = \lambda.
    \end{equation}
    Then,
    \begin{equation}
	q_a(\lambda) = Q_a\pm \frac{ \delta_{ab} P^b } { \sqrt{ 2E - P^2 } } \lambda.
    \end{equation}
    This does not represent the full set of classical solutions. The reason for this is that, when a gauge fixing is performed, the information about the gauge fixing condition itself is lost. This must be the case, otherwise the theory would not be gauge invariant. Thus, these solutions give curves in the space of $q_a$'s, \emph{not} the space of $q_\mu$'s. What is lost is the dynamical information of the gauge fixed variable $q_0$.
    \item {\bf Method 2:} We can parametrize the solutions for $q_a$ in terms of $q_0$, giving
    \begin{equation}
	q_a(q_0) = Q_a \pm \frac{ \delta_{ab} P^b } { \sqrt{ 2E - P^2 } } q_0.
    \end{equation}
    These are indeed the correct classical solutions as they represent straight lines on configuration space with parameters specified by initial conditions. The two branches of the solution represent the ambiguity of specifying an arrow of time, since our formalism is indifferent to the direction in which time is increasing.
\end{itemize}
We can straightforwardly see that method 2 is the correct way of reproducing the classical trajectories. However, this method is \emph{not} compatible with standard techniques used for dealing with gauge systems. This is because gauge invariance requires that the gauge fixed theory is ignorant to the details of the gauge fixing itself. This information, however, is necessary for determining the classical trajectories. Thus, it can not be the case that applying standard gauge theory methods to reparametrization invariant theories will lead to the appropriate quantum theory.

\section{Solving the problem of time: relational quantization}\label{sec:right way}

In the preceding discussion, we have shown how and why standard gauge theory techniques fail to deliver the appropriate quantum theory when applied to theories with global Hamiltonian constraints. According to our diagnosis it is this inappropriateness of the standard canonical quantization techniques that leads to the problem of time.  Our proposed solution to the problem is not to abandon these techniques altogether -- to do so would be to deny ourselves access to number of important mathematical results. Rather, we will outline a formal procedure for modifying an arbitrary globally reparametrization invariant theory such that existing gauge methods \emph{can} be applied and lead to the appropriate quantum theory. In doing so, we will also allow for the construction of a class of quantum theories featuring dynamics with respect to a relational time.

\subsection{Formal procedure}\label{sec:formal procedure}

Consider the general reparametrization invariant theory $\mathcal T$ on the phase space, $\Gamma(q,p)$, with symplectic 2--form, $\Omega = dq \wedge dp$, and Hamiltonian constraint, $\ham$. We assume that all other first class constraints have been gauge fixed according to the procedure outline in Section~\ref{sec:gf stand}. We define the central element, $\epsilon$, of the Poisson algebra as an observable that commutes with all functions on $\Gamma$. As such, $\epsilon$ is a constant of motion. Thus, provided we fix its value by observation, it can be added to the Hamiltonian without affecting the theory: $\ham \to \ham + \epsilon$. In our particle models the central element is simply the total energy of the system and is therefore certainly an observable that can be experimentally fixed. In general relativity, as we will see in Section~\ref{sec:gravSD}, the central element is the cosmological constant.

Now consider the two dimensional symplectic manifold $(F,\Omega_{F})$ coordinatized by $\epsilon$ and its conjugate momentum $\tau$ (i.e. $\Omega_{F}=d\epsilon\wedge d\tau$). We can construct the fibre bundle $(\Gamma_{\text{e}},\Gamma,\pi_{\text{e}},F)$ where $F$ is the fibre, $\Gamma$ is the base space, $\Gamma_{e}$ is the fibre bundle itself, and $\pi_\text{e}$ is a continuous surjection $\pi_{e}:\Gamma_\text{e}\rightarrow \Gamma$. For our purposes, it will be sufficient to consider a trivial bundle structure so that $\Gamma_\text{e}$ is simply the direct product of $\Gamma$ with $F$. The symplectic structure on $\Gamma_\text{e}$ is, thus, given by the non--degenerate symplectic form $\Omega_\text{e} = \Omega \times \Omega_{F}$, which endows $\Gamma_\text{e}$ with a Poisson structure. The overall picture is illustrated in Figure (\ref{fig:fibre}). 
\begin{figure}
 \centering
 \includegraphics[width=.6\textwidth]{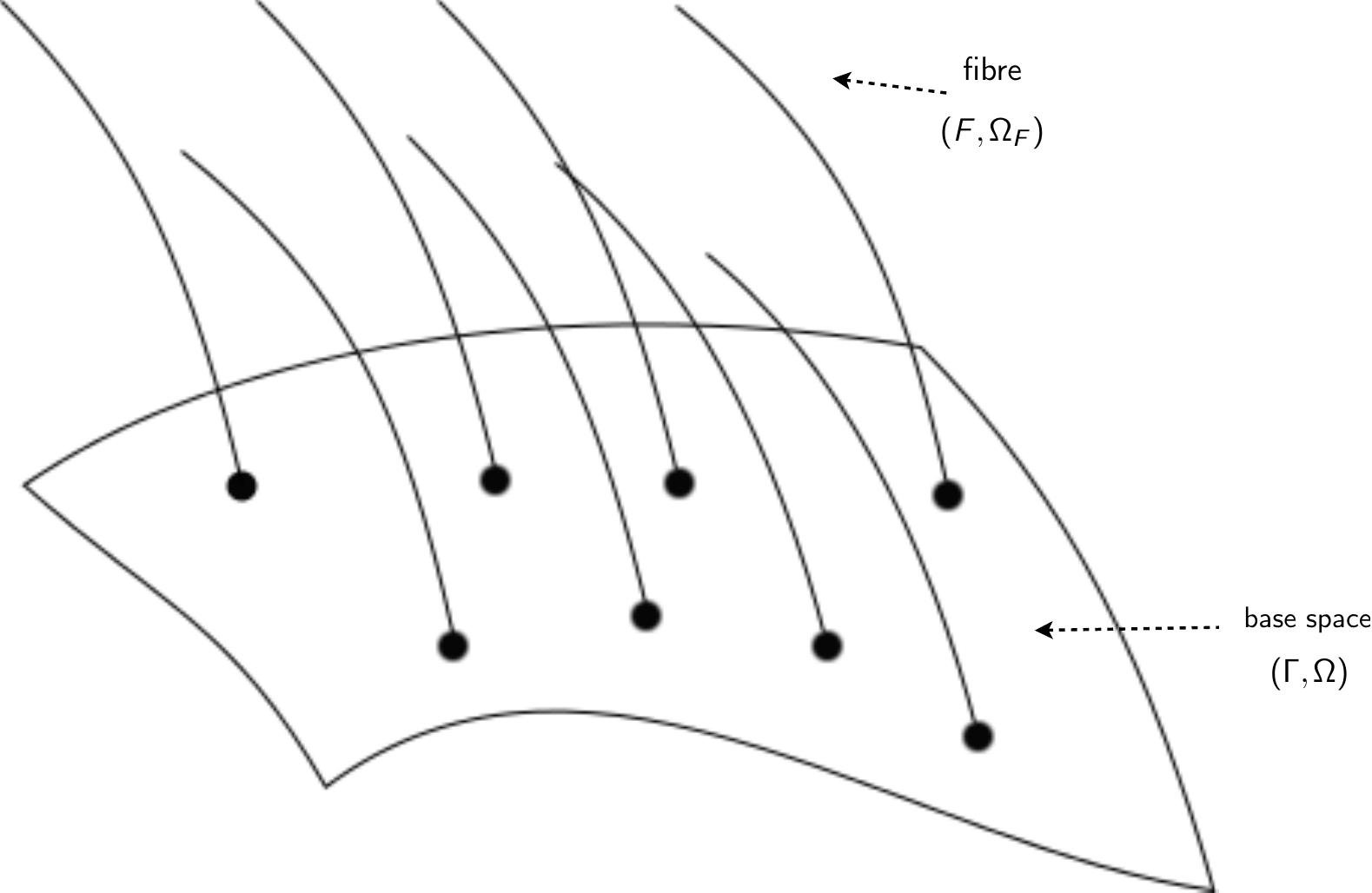}
 \caption{\small{This picture shows the fibre bundle structure of the extended theory. The base space is the phase space of the original theory, $\Gamma(q,p)$, and the fibres are two dimensional symplectic manifolds coordinatized by $\epsilon$ and $\tau$.}}
 \label{fig:fibre}
\end{figure}
We propose that the fibre bundle $\Gamma_\text{e}$ is the phase space, $\Gamma_\text{e}(q,p,\tau,\epsilon)$, of an extended theory $\mathcal T_{\text{e}}$ that, when quantized with conventional gauge theory methods, leads to a quantum theory that: \textbf{a)} correctly describes the classical solutions of $\mathcal T$ in the semi--classical limit without any additional assumptions, and \textbf{b)} describes quantum dynamics with respect to a relational notion of time.

We can establish \textbf{a)} as follows. First, consider $\mathcal T_{\text{e}} = \{\Gamma_\text{e},\Omega_{\text{e}}, \ham_{\text{e}}\}$ where $\ham_{\text{e}}$ is the extended Hamiltonian constraint
\begin{equation}\label{eq:extended Ham}
    \ham_\text{e} = \ham' + \epsilon,
\end{equation}
where $\ham'(q,p, \tau, \epsilon)$ is the pullback of $\ham(p,q)$ under the bundle projection. Next, consider the constraint surface $\Sigma$ defined by $H_{\text{e}}=0$ and the closed degenerate two form $\omega_{\text{e}} \equiv \lf.  \Omega_{\text{e}} \rt\rvert_\Sigma$. The null direction of $\omega_\text{e}$ is generated by the Hamilton vector field $v_{\ham_\text{e}}(\cdot) = \pb{\cdot}{\ham_\text{e}}$, where the Poisson structure on $\Gamma_\text{e}$ is used to compute the brackets. This vector field spans the kernel of $\omega_\text{e}$ and is a codimension 1 submanifold of $\Sigma$. Crucially, the kernel of $\omega_{\text{e}}$ does not, by definition, contain any physically relevant dynamical information because the null directions have a trivial parametrization given by $v(\tau)$ and are, thus, associated with the physically trivial extension procedure. Thus, a gauge fixing (eg., $\tau = \text{const}$) on $\Sigma$ corresponds simply to a section on the bundle. Such a gauge fixing selects a gauge fixed surface $\Sigma_\text{gf}$ that is non--degenerate by construction. It is important to observe that $\Sigma_\text{gf}$ has the same dimension as the original phase space since the constraint $\ham_\text{e}$ and the gauge fixing condition each reduce the phase space degrees of freedom by one, thus eliminating the original auxiliary degrees of freedom $\tau$ and $\epsilon$. Since the reduced phase space is isomorphic to a section on the bundle, it is also isomorphic to the base space. The classical solutions are contained, therefore, in the reduced phase space because the projection $\pi_\text{e}$ maps
\begin{equation}
    \pi_\text{e}: v_{\ham_\text{e}} \to v_\ham,
\end{equation}
which is the Hamilton vector field of the original Hamiltonian. In Section~\ref{sec:solution}, we shall first show how relational quantization is achieved in practice by considering an explicit example and then give a physical interpretation of our result.

\subsection{Proposed solution} \label{sec:solution}
  
We can establish \textbf{b)} as follows. The path integral for this theory is defined using the methods outlined in Section~\ref{sec:gf stand} and using boundary conditions for $\tau$ that are consistent with the value of $\epsilon$ determined observationally. Note that $\ham_\text{e}$ is already in the form $\ham = \hpara + \hperp$. Using $\hpara = \epsilon$ and $\hperp = \ham'$, we can treat $\tau$ as an internal clock by imposing the gauge fixing condition $\tau = \lambda$. Thus, $\tau$ is an ephemeris clock for the theory $\mathcal T$ as it is canonically conjugate to $\ham$ under the bundle projection. The gauge fixed path integral is
\begin{equation}\label{eq:pi good}
    I = \int \mathcal Dq\mathcal D\tau \mathcal Dp\mathcal D\epsilon \delta(\ham_\text{e}) \delta(\tau - \lambda) \det \lf| \pb{\ham_\text{e}}{\tau - \lambda} \rt| \exp \lf\{ i \int d\lambda \lf( \dot q p+\dot \tau \epsilon \rt) \rt\}
\end{equation}
Integration over $\tau$ and $\epsilon$ leads to:
\begin{equation}
    I_{\mathcal{T}} = \int \mathcal Dq \mathcal Dp \exp \lf\{ i \int d\tau \lf[ \frac{dq}{d\tau} p - \ham' \rt] \rt\},
\end{equation}
which obeys the differential equation
\begin{equation}
    i \diby{\Psi}{\tau} = \hat\ham' \Psi.
\end{equation}
Thus, we obtain a time--dependent Schr\"odinger equation where the Hamiltonian is the Hamiltonian of the original theory and the time variable $\tau$ has a classical analogue corresponding to the total change of the system. We have, therefore, passed to a quantum theory where evolution is with respect to a relational notion of time. Although a convenient gauge choice has been made to write this result there is still freedom to use an arbitrary reparametrization $\tau = f(\lambda)$ as the path integral is invariant under the choice of gauge fixing functions. This implies that the fundamental symmetry of the classical theory is still respected quantum mechanically.  

As an example, we can apply this quantization procedure to the toy model of Section~\ref{sec:Jacobi}. The central element $\epsilon$ is identified with the negative of the total energy $E$ of the system. We now extend the phase space to include $\epsilon$ and its conjugate momentum $\tau$ and $E \to -\epsilon$ in the Hamiltonian,
\begin{equation}
    \ham_\text{e} = \frac {\delta_{\mu\nu}}{2} p^\mu p^\nu + V(q) + \epsilon.
\end{equation}
Using the gauge fixing $\tau = \lambda$, the quantum theory is given by the path integral
\begin{equation}\label{eq:pi PPD}
    I = \mathcal Dq_\mu \mathcal Dp^\mu \exp \lf\{ i \int d\tau \lf[ \frac{dq_\mu}{d\tau} p^\mu - \lf( \frac {\delta_{\mu\nu}}{2} p^\mu p^\nu + V(q) \rt) \rt] \rt\}.
\end{equation}
This corresponds to the time--dependent Schr\"odinger theory
\begin{equation} \label{eq:TDSE eg}
    i \diby{\Psi}{\tau} = \lf[ -\diby{^2}{q_\mu^2} + V(q) \rt] \Psi = \hat\ham \Psi.
\end{equation}

In the semi--classical limit, \eq{TDSE eg} reduces to the HJ equation for the phase, $S$, of the wavefunction
\begin{equation}
    \frac {\delta_{\mu\nu}} 2 \diby S {q_\mu}\diby S {q_\nu} + \diby S \tau + V(q) = 0.
\end{equation}
We can do a separation ansatz of the form
\begin{equation}
    S(q, \tau; P, \epsilon) = W(q,P) - E\tau,
\end{equation}
where $W(q,P)$ solves the equation
\begin{equation}
    \frac {\delta_{\mu\nu}} 2 \diby W {q_\mu}\diby W {q_\nu} + V(q) = E.
\end{equation}
We, thus, recover the usual HJ formalism. This procedure, however, is invariant under $\lambda \to f(\lambda)$ so that we maintain the required reparametrization invariance.

In essence, our proposal is that, for a given reparametrization invariant theory with a single Hamiltonian constraint, we can derive the correct quantum theory by applying the standard quantization techniques to an extended version of the original theory. It is important to note that this extended theory is merely an \textit{intermediary formalism}: the relational quantum theory that is derived should be understood as constituting the quantum analogue of the original classical theory and not the extended classical theory.  

For the simplest class of reparametrization invariant models (including our toy model) -- often called \textit{Jacobi's theory} -- relational quantization is equivalent to treating the standard quantization of a parameterized particle model as the quantum analogue to the classical Jacobi's theory -- i..e the parameterized particle model plays the role of the intermediary formalism.\footnote{For an elegant treatment of both Jacobi and parameterized particle models the reader is referred to \cite{Lanczos:1970}} Thus, \textit{mathematically}, the quantum formalism we arrive at is in fact equivalent to that derived (for instance) by Henneaux and Teiteboim \cite{Henneaux:1992a} (again see p.280) when considering the quantization of a parameterized particle theory. However, \textit{physically} our result is importantly different since such authors, following the standard approach, consider the quantum analogue of \textit{Jacobi's theory} to be a Wheeler--DeWitt theory without fundamental temporal structure. As shall be explained in the next section, this difference of interpretation with regard to the correct quantization of simple reparameterization invariant particle models has important implications within the gravitational context.

\section{Gravity and relational quantization} \label{sec:gravSD}

The two principle purposes of the foregoing analysis were to present a non--standard treatment of the origin of the problem of time in quantum gravity and, in that light, to offer a new proposal for a possible solution. Despite certain major structural similarities, the simple reparametrization invariant particle models we have been considering are, in some respects, clearly \textit{disanalogous} to the full theory of  general relativity. Most importantly, in its ADM formulation \cite{Arnowitt:1960,Arnowitt}, GR has a Hamiltonian constraint that is local in space and is thus an infinite set of constraint functions. The form of and relationship between these constraints indicate that, rather than merely being reparametrization invariant, ADM GR has the larger and more subtle symmetry of \textit{foliation invariance} \cite{Teitelboim:1973}. Although we may expect key elements of our arguments against Hamiltonian constraints as gauge generating to carry over to the gravitational case,\footnote{In particular, in GR it is still the case that standard gauge theory methods lead to a classical theory without even minimal temporal structure. See \S3 of \cite{Thebault:2011c}.} we \textit{may not} reasonably expect our relational quantization procedure to be applicable to the full gravitational case as it stands: its consistency would seem to depend upon the theory in question having a single global Hamiltonian constraint. However, it has recently been shown to be possible to reformulate gravity as a theory with a single global Hamiltonian constraint by \emph{trading} foliation invariance for local conformal invariance \cite{gryb:shape_dyn,Gomes:linking_paper}. This reformulation of gravity, called Shape Dynamics (SD), is dynamically equivalent to pure ADM gravity on compact spatial Cauchy surfaces and is generally equivalent to ADM gravity when certain natural gauge fixing conditions are met.

In this section, we will first apply our relational quantization procedure to SD and see that we get a quantum theory that is analogous to the Dirac quantization of unimodular gravity \cite{brown:gr_time,Henneaux_Teit:unimodular_grav,Unruh:unimodular_grav,Unruh_Wald:unimodular}. We next apply our relational quantization procedure to GR, ignoring the issues associated to the local nature of the Hamiltonian constraint, and find that, indeed, we get the Dirac quantization of unimodular gravity. Finally, we will comment on the significance of unimodular SD as a possible solution to the problem of time in quantum gravity.

\subsection{Relational quantization of Shape Dynamics}

First, we begin by quickly reviewing the key elements of shape dynamics (for a more detailed exposition we recommend \cite{JuliansReview} for conceptual motivations, \cite{gryb:thesis} for a more technical introduction, and \cite{Gomes:linking_paper} for a formal development).

SD is a Hamiltonian theory of evolving spatial geometry. There is a single global `time' variable that labels ordered sequences of geometries and their associated conjugate momenta. These continuous sequences trace out the solutions of SD and can be used to construct a manifold with topology $\mathbb R \times \Sigma$, where $\mathbb R$ is the topology of the time parameter and $\Sigma$ is the topology of the spatial manifold.\footnote{For simplicity, we will assume that $\Sigma$ is compact without boundary.} The phase space is, thus, constructed from 3--dimensional metrics $g_{ab}$ and their conjugate momentum densities $\pi^{ab}$ (where $a = 1\hdots 3$). This is identical to the phase space used by ADM to construct a Hamiltonian formulation of GR. However, unlike GR, the time parameter is global and not a local function of the spatial manifold. This means that, in general, it is not possible to construct a 4--dimensional metric invariant under arbitrary 4--diffeomorphisms from the evolution of the 3--dimensional variables used in the theory. In GR, this is possible because of the precise form of the constraint algebra of the theory, which is very different in SD. Despite this difference in the constraint algebra, it is possible under certain conditions,\footnote{The solutions must admit at least one foliation where the spatial Cauchy hypersurfaces have constant mean (extrinsic) curvature -- i.e. be `CMC foliable.' One may argue that all physically reasonable solutions will satisfy this condition which, in any case, is not dramatically stronger than the global hyperbolicity assumption that is fundamental to ADM GR.} to prove that the dynamical solutions of SD are gauge equivalent to those of GR \cite{gryb:shape_dyn,Gomes:linking_paper}. Thus, shape dynamics and general relativity can be considered to be physically equivalent theories.

The Hamiltonian of SD is given by the sum of three first class constraints $H_\text{gl}$, $H_{\text{diff }a}$, and $H_\text{conf}$ with associated Lagrange multipliers $N(t)$, $N^a(x,t)$, and $\rho(x,t)$ respectively
\begin{equation}
    H_\text{SD} = N(t) H_\text{gl} + \int d^3 x \lf[ N^a(x,t) H_{\text{diff }a} + \rho(x,t) H_\text{conf} \rt].
\end{equation}
Note that the \emph{lapse} $N(t)$ is always homogeneous because the time variable is global. These constraints can be split into two kinds: i) the constraints that generate gauge transformations and have associated symmetries and ii) the constraint that generates the dynamics. The constraints $H_\text{diff}$ and $H_\text{conf}$ are linear in the momenta and fall under the first kind. The explicit form of these constraints will not be necessary but we can understand the significance of each by noting the gauge symmetries that they generate. The \emph{diffeomorphism} constraint $H_\text{diff}$ is common to both SD and GR. It generates infinitesimal spatial diffeomorphisms and requires that only the geometric information contained in the metric, and not the coordinate information, be physical. The \emph{conformal} constraint, $H_\text{conf}$, generates conformal transformations of the metric of the form
\begin{equation}
    g_{ab} \to e^{\phi} g_{ab}.
\end{equation}
These conformal transformations, however, have a global restriction that the total volume of space be preserved. Physically, $H_\text{conf}$ requires that the information about the local scale is unphysical. Thus, only angles and ratios of lengths are observable. However, the global scale, set by the spatial volume of the universe, is \emph{not} gauge. This global restriction on scale invariance is crucial because it allows $H_\text{conf}$ to be first class with respect to the non--trivial global constraint $H_\text{gl}$. In terms of the number of degrees of freedom, this global restriction is also necessary because the two phase space degrees of freedom killed by $H_\text{gl}$ are recovered by imposing this restriction on $H_\text{conf}$. Thus, the total number of constraints is still equal to that of GR.

The dynamics are generated by the global Hamiltonian constraint $H_\text{gl}$. This constraint is \emph{uniquely} defined by the two requirements: i) that the classical dynamics and initial value problem of SD are identical to that of GR and ii) that it be first class with respect to $H_\text{diff}$ and $H_\text{conf}$. It is important to point out that the first class requirement implies that $H_\text{gl}$ is invariant under both spatial diffeomorphisms and conformal transformations that preserve the volume. Unfortunately, $H_\text{gl}$ is non--local in the sense that it is defined through the formal solution of an elliptic differential equation (given explicitly in \cite{gryb:gravity_cft}) which is a modified version of the so--called Lichnerowicz--York \cite{York:york_method_prl} equation. It can however, be given explicitly in terms of different perturbative expansions. For our purposes, we will only need the first term of $H_\text{gl}$ in a large volume expansion. This is a well defined expansion in SD because the volume is a gauge invariant quantity. The details can be found in \cite{gryb:gravity_cft,gryb:thesis}. We will only quote the result:
\begin{equation}
    H_\text{gl} = 2\Lambda - \frac 3 8 P^2 + \mathcal O(V^{1/3}),\label{eq:SD large V}
\end{equation}
where $\Lambda$ is the cosmological constant and $P$ is proportional to the mean of the trace, $\pi^{ab} g_{ab}$, of the metric momenta. For completeness we include its definition (although it will not be used):
\begin{equation}
    P = \frac 2 3 \frac 1{\int d^3 x \sqrt g} \int d^3x \pi^{ab} g_{ab}.
\end{equation}
Physically, it is helpful to note that $P$ is the variable canonically conjugate to the spatial volume and is equal to the York time, which is always homogeneous in SD. Note that, to this order in $V$, the Hamiltonian is homogeneous and leads to the Friedmann universe with pure cosmological constant. Also, in this limit gravity is equivalent to a free particle model like the ones treated earlier in the text, justifying the their use as valid toy models for quantum gravity.

We have now laid out sufficient structure to perform our quantization procedure on SD. We define the central element of the observable algebra, $\ep$, through the Poisson bracket relations
\begin{equation}
    \pb{\ep}{g_{ab}} = \pb \ep {\pi^{ab}} = 0.
\end{equation}
Its conjugate momentum, $\tau$, is defined by $\pb \tau \ep = 0$. We extend the classical phase space to include $\tau$ and $\ep$ with the Poisson brackets given above and extend the classical Hamiltonian constraint
\begin{align}
    H_\text{gl} \to& \ep + H_\text{gl} \\
                 =& \ep + 2\Lambda - \frac 3 8 P^2 + \mathcal O(V^{1/3}).\label{eq:uni SD}
\end{align}
That this produces an equivalent classical theory can be seen by computing the classical equations of motion for $\ep$
\begin{equation}
    \dot\ep = \pb \ep {N H_\text{gl}} = 0.
\end{equation}
Thus, $\ep$ is a constant of motion. We can then integrate out $\ep$ in the classical theory and obtain a new Hamiltonian that is just shifted from the original by the constant of motion, $\mathcal E$, associated to $\ep$. Clearly, the extension procedure has the effect of redefining the cosmological constant
\begin{equation}
    \Lambda \to \Lambda + \frac 1 2 \mathcal E.
\end{equation}
From an operational point of view this requires a change of philosophy: the cosmological constant is seen as a constant of motion rather than a constant of Nature. However, this new interpretation has no effect on the physical predictions of the classical theory.

Despite the fact that the classical theory is unaltered, the quantum theory is noticeably different from that obtained by Dirac quantization because we require that $\ep$ be promoted to an operator. This leads to the following operator constraints on the SD wavefunctional, $\Psi$,
\begin{equation}
    \hat\ep \Psi = -i \diby{\Psi}{\tau} = \lf( 2\Lambda - \frac 3 8 \hat P^2 + \hat H_{\mathcal O(V^{1/3})} \rt) \Psi.
\end{equation}
The cosmological constant can be removed by simply shifting the eigenvalues of the $\hat\ep$ operator, just as in the classical theory. We see that the theory we obtain is equivalent to that obtained if we treated the cosmological constant as a global canonical variable rather than a coupling constant. We get a definite time evolution in terms of the global parameter $\tau$.

We can better understand the meaning of this relational quantum theory by considering the nature of the classical intermediary formalism associated with the extend Hamiltonian constraint (\ref{eq:uni SD}). This can be seed to be the SD equivalent to the unimodular gravity theory developed in \cite{brown:gr_time,Henneaux_Teit:unimodular_grav,Unruh:unimodular_grav,Unruh_Wald:unimodular}. In particular, in \cite{brown:gr_time}, it is shown that promoting the cosmological constant to a canonical variable, in the context of GR, produces a time--dependent quantum theory where the time variable, $\tau$, is canonically conjugate to the cosmological constant. In this case, as in ours, $\tau$ is interpreted as the 4--volume of the universe. In GR, the situation is a bit more subtle than in SD because $\ep$ is allowed to vary over space. However, as is shown in detail in \cite{Henneaux_Teit:unimodular_grav}, there is a secondary constraint $\nabla_a \ep = 0$ that enforces the homogeneity of the $\ep$. Once this constraint is enforced, it is straightforward to see that the Hamiltonian obtained in \cite{Henneaux_Teit:unimodular_grav} is equivalent to the modified SD Hamiltonian \eq{uni SD}. Thus the relational quantization of shape dynamics leads to a formalism equivalent to the Dirac quantization of unimodular shape dynamics. We might,  therefore, expect that a prospective relational quantization of ADM GR would be equivalent to a Dirac quantization of unimodular gravity. This possibility will be investigated in the following section.

\subsection{Relational quantization of general relativity}

The relational quantization procedure presented in this paper was motivated by what happens in reparametrization invariant theories where a single Hamiltonian constraint generates the dynamics. Although the situation is more subtle in GR, where there is a different Hamiltonian constraint for each spatial point, it may still be constructive to check what happens when we apply our quantization procedure in this case. Before doing so, we must briefly review the general structure of GR in Hamiltonian form. We will only give the information necessary to present our result. For more details on the Hamiltonian formulation of GR, see \cite{Arnowitt:1960,Arnowitt,Thiemann:2007}.

The GR Hamiltonian is defined on the same phase space as SD and is the sum of two local constraints $S$ and $H_{\text{diff }a}$ with associated Lagrange multipliers $N(x,t)$, $N^a(x,t)$
\begin{equation}
    H_\text{ADM} = \int d^3x \sqrt g \lf( N S + N^a H_{\text{diff }a} \rt).
\end{equation}
The diffeomorphism constraint $H_{\text{diff }a}$ is identical to that of SD. The ADM Hamiltonian constraint $S$, which generates dynamics, is a local function of space and is quadratic in the momenta. The explicit form of the constraints and the constraint algebra will not be necessary for our discussion.

To perform the relational quantization, we must introduce the central element of the observable algebra $\ep$. However, because the Hamiltonian constraint, $S$, is a local function of space, so too must be $\ep$. Thus, we must shift $S$ in the following way
\begin{equation}
    S(x,t) \to S(x,t) + \ep(x,t),
\end{equation}
where we still have
\begin{equation}
    \pb{\ep}{g_{ab}} = \pb \ep {\pi^{ab}} = 0.
\end{equation}
The time variable, $\tau(x,t)$, canonically conjugate to $\ep(x,t)$ must also be a local function of space. It would seem that this would produce a qualitatively different theory form the unimodular one previously considered. However, this exact theory has been treated in detail in \cite{sg:dirac_algebra}. In Section~3.2.4 of that paper, it is shown that the consistency of this theory requires a secondary constraint of of the form $\nabla_a \ep = 0$ and that the resulting theory is identical to the unimodular theory given in \cite{Henneaux_Teit:unimodular_grav}. Thus, the quantization procedure presented in this paper applied to GR leads to the standard Dirac quantization of unimodular gravity.

\subsection{Comments on unimodular shape dynamics}

Unimodular gravity has been proposed as a possible solution to the problem of time \cite{Sorkin:forks_unimodular}. The homogeneous and isotropic case (corresponding to the large volume limit given in Equation~\eq{SD large V}) has been studied and unitary solutions have been found to exist \cite{sorkin:unmodular_cosmology}. Furthermore, it has been argued that treating the cosmological constant as an integration constant rather than a coupling constant could provide a resolution of the cosmological constant problem \cite{Smolin:unimodular_grav}. Despite these hopes, there are well--known criticisms for treating unimodular gravity as a genuine solution to the problem of time. These are summarised in \cite{Kuchar:unimodular_grav_critique}. The essential argument is that foliation invariance in GR makes it impossible to genuinely define a global time, which is necessary in the unimodular description. We see these difficulties, in our context, as arising from the fact that our quantization procedure was designed only to work for theories with a global Hamiltonian. As a result, we can not claim to resolve these difficulties in the context of GR. However, in SD, the situation is considerably improved. In this case, there is a genuine global time parameter and a single Hamiltonian constraint generating dynamics. Thus, the unimodular SD theory presented above is free from the criticisms presented in \cite{Kuchar:unimodular_grav_critique} and provides a proposal for a genuine solution to the problem of time in quantum gravity. 

In essence, our solution is constituted by the application of a three stage procedure: i) translate ADM GR into equivalent shape dynamics formalism; ii) apply extension procedure to construct unimodular shape dynamics; iii) apply standard Dirac quantization to derive dynamical theory of quantum gravity. Of these three steps, the basis behind the first is perhaps the most contestable; does moving to the shape dynamics formalism not simply amount to sweeping the problem of foliation invariance `under the rug', rather than solving it? We think not. \textit{On the one hand}, if one considers shape dynamics a fundamental theory of gravity, then we have moved to a formalism that makes manifest a physical deep symmetry triplet of reparametrization invariance, three dimensional diffeomorphism  invariance and three dimensional scale invariance. From this perspective, the issue of retaining foliation invariance within quantum gravity is simply no longer relevant. \textit{On the other hand}, if one insists that general relativity should retain its fundamental status, then -- due to duality between that theory and shape dynamics -- one can still consider the procedure i-iii above as providing a potential methodology to explore the phenomenology of foliation invariant quantum gravity. In either case, a quantum theory of unimodular shape dynamics offers an interesting new possibly within the theory space of quantum gravity and warrants consideration of its explicit details, formal consistency and potential for  application. Such an investigation will be the subject of future work. 

\section{Closing remarks}

Standard quantum mechanics inherits from Newtonian theory a bipartite absolute temporal structure: both the ordering of events (i.e. topological structure) and the duration (i.e. metric structure) are externally fixed. Quantum systems based on a Wheeler--DeWitt--type equation completely do away with both of these aspects of time. Our relational quantization procedure allows us to dispense with an absolute notion of duration whilst retaining the topological structure necessary to recover the classical dynamics. Although it may be possible to recover temporal phenomenology from a timeless formalism, since `the Wheeler--DeWitt equation does not know about...ordering parameter[s]' (\cite{Halliwell:emergent_time} p.2) the recovery of the full classical notion of time is not possible without some additional principle (e.g. Barbour's time capsules \cite{barbour:eot}). Our proposal does not require any additional principle since the minimum temporal structure is retained throughout.

The quantization procedure presented here constitutes a novel solution to the problem of time for globally reparametrization invariant theories. It provides a fundamental insight into the treatment of the Hamiltonian constraint in quantum gravity, the definition of observables, and the recovery of dynamics. Additionally, given that equitable duration emerges as a purely classical notion, our analsysis may provide insight into the interpretation of classical observations in quantum theory. Relational quantization has immediate applicability to shape dynamics and, in that context, leads to a formalism (closely related to unimodular gravity) that can be understood as a dynamical theory of quantum gravity.   

\section*{Acknowledgements}

We would like to thank Julian Barbour for helpful comments and criticisms and Lee Smolin for encouraging us to look more closely at the Hamilton--Jacobi formalism. We would also like to thank the participants of the 2$^\text{nd}$ PIAF conference in Brisbane (in particular Hans Westman) for their inspiring questions on the problem of time. Finally, we would like to extend a special thanks to Tim Koslowski for many useful discussions and for making valuable contributions to our overall understanding of this work. Research at the Perimeter Institute is supported in part by the Government of Canada through NSERC and by the Province of Ontario through MEDT.

\bibliographystyle{utphys}
\bibliography{mach,Masterbib}

\end{document}